\documentclass[12pt,preprint]{aastex}
\usepackage{psfig}
\newcommand\fsec{\mbox{$\rm.\mkern-4mu^s$}}%

\shortauthors{Stecklum et al.}
\shorttitle{Discovery of a circumstellar disk in the Bok globule CB\,26}

\slugcomment{To appear in The Astrophysical Journal}

\begin{document}

\newcommand\farcd{\mbox{$.\!\!^\circ$}}%

\title{High-resolution NIR Observations of the Circumstellar Disk System in the Bok Globule CB\,26 $^1$}

\author{
B. Stecklum\altaffilmark{2},
R. Launhardt\altaffilmark{3},
O. Fischer\altaffilmark{4},
A. Henden\altaffilmark{5},
Ch. Leinert\altaffilmark{3},
H. Meusinger\altaffilmark{2}}

\altaffiltext{1}{Based on observations carried out at the German Spain Astronomical
   Center Calar Alto, Spain, the Th\"uringer Landessternwarte Tautenburg, Germany,
   and the U.S. Naval Observatory, Flagstaff Station.}

\altaffiltext{2}{Th\"uringer Landessternwarte Tautenburg, Sternwarte 5, D-07778 Tautenburg,
   Germany, stecklum@tls-tautenburg.de}

\altaffiltext{3}{MPI f\"ur Astronomie, Heidelberg, K\"onigstuhl 17, D-69117 Heidelberg, Germany}

\altaffiltext{4}{AG Didaktik der Physik und Astronomie, Friedrich-Schiller-Universit\"at Jena,
   Max-Wien-Platz 1, D-07743 Jena, Germany}

\altaffiltext{5}{Universities Space Research Association/US Naval Observatory,
   Flagstaff Station, P.O. Box 1149, Flagstaff, AZ 86002-1149 USA}

\begin{abstract}

We report on results of near-infrared and optical observations of the mm disk
embedded in the Bok globule CB\,26 \citep{lasa01}.
The near-infrared images show a bipolar reflection nebula %
with a central extinction lane which coincides with the %
mm disk. 
Imaging polarimetry of this object yielded a polarization pattern which is
typical for a young stellar object surrounded by a large circumstellar disk
and an envelope, seen almost edge-on. 
The strong linear polarization in the bipolar lobes is caused by single
scattering at dust grains and allowed to locate the illuminating source which
coincides with the center of the mm disk.
 The spectral energy distribution of the YSO embedded in CB\,26 resembles
that of a Class~I source with a luminosity of 0.5\,L$_\sun$.
Using the pre-main-sequence evolutionary tracks and the stellar mass inferred
from the rotation curve of the disk, we derive an age of the system of
$\le$\,10$^6$\,yr.
H$\alpha$ and [SII] narrow-band imaging as well as optical spectroscopy revealed
an Herbig-Haro object $6\farcm15$ northwest
of CB\,26~YSO\,1, perfectly aligned with the symmetry axis of the
bipolar nebula. This Herbig-Haro object (HH\,494) indicates
ongoing accretion and outflow activity in CB\,26~YSO\,1. Its excitation
characteristics
indicate that the Herbig-Haro flow is propagating
into a low-density environment.
We suggest that CB\,26~YSO\,1 represents the transition stage between embedded 
protostellar accretion disks and more evolved protoplanetary disks around T
Tauri stars in an undisturbed environment. 
\end{abstract}


\section{Introduction}

Bok globules are small, opaque, and isolated nearby molecular clouds. Their cores
usually have a quite simple structure which makes them ideally suited for studies
of isolated low-mass star-formation \citep{bok77,leun85}.
CB\,26 \citep{clba88} was classified by \citet{clem91} as quiescent, with
little or no star-forming activity.
The object was included in a comprehensive investigation of Bok globules using
1.3\,mm dust continuum and CS molecular line observations \citep{laun97,laun98} 
as well as near-infrared (NIR) imaging (Launhardt et al., in prep.).
\citet{laun97} found an unresolved 1.3\,mm source with F$_{1.3mm}=160\pm14$\,mJy
slightly offset, but within the positional error ellipse from of IRAS\,04559+5200
and concluded that a protostellar core is present in CB\,26.
Recent interferometric observations by \citet{lasa01} revealed that the major 
fraction of the thermal dust emission at mm wavelengths is due to a young,
edge-on seen protoplanetary disk of diameter $\sim 400$\,AU and mass $\rm \sim 0.1\,M_\sun$.
They also suggest that CB\,26 is part of the Taurus/Auriga 
complex at the distance of 140\,pc and thus nearer than claimed by \cite{laun97}.
The revised distance will be adopted in the present paper.

Near-infrared (NIR) imaging of globules led to the detection of young stellar objects
(YSOs) in many of them which are often associated with infrared reflection nebulae. 
One such example is CB\,230~IRS which has a cometary shape due to light scattered in the
northern (blueshifted) lobe of the outflow \citep{yun94, lau01}. In such cases, the strong
extinction even at NIR wavelengths prevents the direct view on the embedded pre-main 
sequence (PMS) star. However, it can be located using imaging polarimetry \citep{scar91}
since in the case of a single illuminating source, the polarization vector of
single-scattered photons is perpendicular to the radius vector.
The polarization pattern of YSOs is indicative for the dust properties and can often be
explained by scattering in a circumstellar disk/envelope with bipolar cavities
\citep{els78, fisch94}.

Our NIR imaging and polarimetry observations of CB\,26 revealed the presence of an
infrared source which is associated with the mm disk.
In the following, we will refer to the embedded source in the globule 
as CB\,26\,YSO\,1. This includes the NIR reflection nebula (called  CB\,26\,IRS\,1),
the mm disk, and the central star.

\section{Observations and data reduction}  \label{sec-intro}

J, H, and Ks images were obtained using the MAGIC camera \citep{herb93} at the
3.5-m telescope of the Calar Alto observatory in 1994 January at the pixel scale 
of 0\farcs32. The seeing was $\sim1\arcsec$. The total field of view (FoV) of the 
mini mosaics is about 2\arcmin.
The total integration time of 800\,s for each band led
to 3$\sigma$\ surface brightness detection limits of 21.2, 20.5, and 20.1
mag/$\Box$\arcsec. Astrometry was established using 5 stars in common with the 
Digitized Sky Survey 2 (DSS-II), yielding an accuracy of $\sim0\farcs2$. 
Several photometric standard stars where observed during the night. Final flux densities 
listed in Table \ref{tab1} where derived after background subtraction within 
synthetic circular apertures encompassing the entire NIR nebula.

CB\,26 was re-observed in 1995 March using MAGIC together with a wire-grid polarizer 
at the pixel scale of 0\farcs07 in the K band during sub-arcsecond seeing.
The linear polarization (degree p and angle $\Theta$)
was derived from image stacks taken at four polarizer settings (position
angles: 45, 90, 135, and 180\arcdeg). Images at orthogonal directions
were taken subsequently to minimize the effects of
seeing and transmission variations. The total integration time
per position angle was 200\,s. The images were rebinned to
0\farcs28 in order to gain signal-to-noise ratio. The K-band image shown in Fig.\ \ref{fig-k}
is the sum of the four polarization frames. Its astrometry is based on our previous NIR
images. Instrumental polarization was checked using standard stars from the UKIRT list and 
was found to be negligible compared to the high polarization degrees of CB\,26 reported below.

Additional NIR imaging (H, K, L') was carried out using the 1.55-m Kaj Strand Astrometric 
Reflector of the USNO Flagstaff station together with ASTROCAM utilizing an ALADDIN 
1024$\times$1024 InSb array at the pixel scale of 0\farcs35 in December 2000. 
Total integration times were
720\,, 720\,, and 180\,s, respectively. The NIR images were processed
using sky flats as well as bad pixel removal, and lastly mosaicked. HD\,1160 served as photometric
standard and was observed immediately before the science target at about the same airmass.
Flux densities were derived in the same way as for the MAGIC images.
Optical images (I, H$\alpha$) were taken with the 2048$\times$2048 CCD prime focus camera of the
2-m Tautenburg telescope (diameter of the Schmidt correction plate 1.34\,m) in February 2001 
at the pixel scale of 1\farcs24.
Two exposures were done in each band with integration times of 180\,s and 600\,s,
respectively, to enable the reliable removal of cosmic rays. The images were flat-fielded
with dome flats and astrometrically calibrated using 8 stars from the DSS-II. Their spatial
resolution is $\sim$4\arcsec. Due to the
larger bandwidth, the I-band image is 1.5\, mag deeper than the H$\alpha$ frame.
Since stars with known magnitudes are saturated on the I-band image, we established its
photometry as follows. Ten stars from the USNO2 catalog \citep{mone96} were selected which are
in the unobscured neighborhood of the globule. Their I magnitudes were predicted using the (B--R)
color as a guide for the spectral type. We found a good correlation between the predicted and the
instrumental magnitudes, with a standard deviation of 0.3\,mag sufficient for our purpose.
Additional narrow-band imaging using the [SII]\,6716/6731\AA{} filter with an exposure time
of 2$\times$1200\,s was obtained in September 2001.

Spectroscopy was performed in October 2001  with 
the newly commissioned Nasmyth spectrograph at the 2-m Tautenburg telescope using a total integration
time of 1 hour and a slit width of 2\arcsec, yielding a resolution of $\lambda/\delta\lambda\sim$960. 
The wavelength calibration is based on
night-sky lines. For this purpose, a template spectrum was generated by folding the Osterbrock Sky
Spectrum\footnote{Preparation of the Osterbrock Sky Spectrum files was supported by grant No. ATM-9714636
from the NSF CEDAR program.} (www.nvao.org/NVAO/download/Oster\-brock.html) to the observed resolution.
The dispersion curve
was derived by fitting a second order polynomial to the results of Gaussian fits to 14 lines in
the wavelength region from 6200\,\AA\ to 7100\,\AA\ in both the observed and template spectrum. The
zero point of this calibration was checked using the almost unblended [OI]\,6300\AA\ sky line
which did not deviate by more than 5\,km/s from the zero radial velocity which represents the
1\,$\sigma$ uncertainty of the wavelength calibration.

 An additional Nasmyth spectrum was obtained during good observing conditions in February 2004
using a 1\arcsec{} slit ($\lambda/\delta\lambda\sim$1800) and an integration time of 20 minutes. During this run, the error
of the wavelength calibration was checked by observing the brightest knot of the Herbig-Haro object
(HHO) 366 for which \cite{ball96} found $v_{LSR}=+45$\,km/s. Our estimate of $v_{LSR}=+42\pm3$\,km/s
agrees with their value and confirms the accuracy of the radial velocity estimation based on sky lines.

\section{Results}  \label{sec-res}
Our NIR images  reveal an infrared source (IRS\,1) close to the southwest rim
of the small cometary-shaped Bok globule CB\,26 (Figs. \ref{fig-j} and \ref{fig-col}).
Its morphology resembles that of bipolar reflection nebula. 
The central part of the nebula is bisected by a dark extinction
lane which is seen in the high-resolution NIR images (Figs. \ref{fig-j} and \ref{fig-k}),
but becomes most evident in the true color image shown in Fig. \ref{fig-col}.
The extinction lane is inclined at the position angle (P.A.) $55\pm7$\arcdeg,
which is in good agreement with the P.A. of the mm emission from the circumstellar disk
($60\pm5$\arcdeg; Launhardt \& Sargent 2001).
A more diffuse and redder extension of the nebula stretching to the north-west is separated 
from the central part of the nebula by an additional, more diffuse extinction lane
(Fig.\ \ref{fig-col}, see also Sect. \ref{sec-dis-morph}).
The symmetry axis of the reflection nebula was derived from a weighted linear 
regression to the J-band brightness distribution  clipped at the 3\,$\sigma$ level.
It is oriented along P.A.\,$=144\pm1$\arcdeg, and is thus perpendicular to the plane of the
disk and extinction lane. Since all three features, the reflection nebula, the extinction lane,
and the mm disk are aligned within the small uncertainties, we assume a common symmetry axis
along P.A.\,\,$=145\pm5$\arcdeg.

Figure \ref{fig-k} shows our highest-resolution K-band image and the map of the linear polarization
vectors. Single scattering yields high polarization degrees with the polarization vector perpendicular
to the line connecting the illuminating source and the scattering grain. The most likely
location of the illuminator was therefore derived by minimizing the mean square scalar product
between polarization vectors with $p\ge10$\%\ and their corresponding radius vectors.
This position is very close but not fully coincident with the center of the extinction lane.
It is shifted by 0\farcs3\ to the north-west. However, the formal 1\,$\sigma$\ positional uncertainty
of the center of illumination amounts to 0\farcs7\ (see Fig. \ref{fig-k}) so that this displacement
is not significant.

Area-integrated flux densities of the NIR reflection nebula were derived from aperture photometry
in the calibrated images. Comparison of the H and K magnitudes from MAGIC and ASTROCAM images
showed that they are consistent within $\sim$0.1\,mag. Because of the better signal-to-noise ratio,
we use the MAGIC fluxes. The uncertainty of the L' flux is 0.5\,mag.
NIR flux densities of CB\,26~YSO\,1 and their errors, synthetic aperture diameters as well as 
beam sizes (stellar FWHMs due to seeing) are compiled in Tab. \ref{tab1}.

Our NIR photometry, the IRAS PSC as well as the submm/mm fluxes from \citet{laun97},
\citet{henn01} and Launhardt et al. (in prep.) were used to establish the spectral energy 
distribution (SED) of CB\,26~YSO\,1, shown in Fig. \ref{fig-sed}.
The FIR to mm part of the SED was fitted by modeling the disk
(parameters from Launhardt \& Sargent 2001) plus a two-component
spherical envelope consisting of cold (16\,K) and warm (45\,K) dust
with a mass ratio of 750:1. A dust opacity $\kappa_{\nu}(1.3\,\rm mm)$
of 1\,cm$^2$g$^{-1}$\ together with a frequency dependence
$\kappa_{\nu}\propto \nu^{1.8}$\ was used for the envelope
(Ossenkopf \& Henning 1994).
The best fit to the SED yields a total mass for the extended and 
slightly asymmetric envelope (see Henning et al. 2001) of 
$\rm 0.1\,M_\sun$, which is comparable to the mass of the embedded 
disk (Launhardt \& Sargent 2001). From the submm maps we estimate an
average K-band extinction through the envelope of order 1 mag.
The NIR section of the SED was modeled by a modified 3000\,K black body
(typical for an M star), using an extinction curve
from \citet{riek85} and scattering efficiencies from \citet{kim94}.
No physical parameters were derived from
the simple NIR fit. The mid-infrared part of the SED is very uncertain
and not well-constrained by observations.
Integration over 4\,$\pi$\ yields a luminosity $L_{\rm bol}$\
of CB\,26\,YSO\,1 of $0.47\pm 0.04$\,L$_{\odot}$, of which $\sim$5\%
are from the NIR reflection nebula (i.e., due to emission/scattering at 
$\lambda < 12$\,$\mu$m).

Figure \ref{fig-i} shows the wide-field I-band image of the region around CB\,26.
Overlayed are the contours of the continuum-subtracted H$\alpha$\ image. 
An extended emission line feature is obvious 6\farcm15 northwest of CB\,26~IRS\,1 which is
also present on the [SII] image but lacks any continuum counterpart on the deeper
I-band image. Its equatorial coordinates (J2000) are $\rm RA=04^h59^m29\fsec4,
DEC=+52\degr09\arcmin52\arcsec$.  The inspection of this region on the corresponding
DSS-II-F plate digitized using the MAMA facility \citep{1992doss.conf..103G} shows a faint non-stellar
object at the same location which is missing on the DSS-II-J image. Since the spectral response
of the F plates almost peaks at H$\alpha$ while the J plates are not sensitive to this line,
we conclude that it is the emission line object seen in our narrrow-band frames.

The spectrum of this source is shown in Fig. \ref{fig-spec}.
Strong emission lines (H$\alpha$, H$\beta$, [OI]\,6300\,\AA{}, [NII]\,6584\,\AA{}, [SII]\,6717\AA{},
and [SII]\,6731\,\AA{}) confirm its Herbig-Haro nature.
This newly discovered HHO has the entry 494 in the
catalog of \citep{reip99}.
The symmetry axis of the NIR reflection nebula and the mm disk
(P.A. $145\pm$5\arcdeg) precisely points to the HHO 
(P.A. of the connecting line 147\farcd5), suggesting a physical connection between the two 
(see Fig. \ref{fig-i}).
The almost edge-on view of the circumstellar disk (Launhardt \& Sargent 2001;
see also Sect. \ref{sec-dis-morph}) implies that the projected linear separation of 0.25\,pc
($\approx 5\times 10^4$\,AU) between
CB\,26~YSO\,1 and HHO\,494 is very close to the true separation.
The radial velocity of HHO\,494 was derived from the lines shown in Fig.
\ref{fig-spec} and amounts to $v_{LSR}^{HHO}=-45\pm7$\,km/s.

\section{Discussion}  \label{sec-dis}

\subsection{Morphology of the disk and envelope} \label{sec-dis-morph}

The morphology and SED of the  embedded YSO resembles that of low-mass Class\,I source consisting 
of a PMS star surrounded by a circumstellar disk embedded in an envelope
\citep{1999AJ....117.1490P,1999A&A...352L..73Z}.
While the embedded YSO is not seen directly due to high extinction in the surrounding disk,
stellar light scattered at dust grains in the envelope below and above the disk becomes 
evident as bipolar NIR reflection nebula. 

The observed centro-symmetric polarization pattern (Fig. \ref{fig-k}) is caused by light scattering in
this bipolar geometry, resulting in very high values of $\rm p\sim50\dots90$\% at the outer lobes
due to single scattering. The linearly aligned polarization vectors close to the disk might be
caused by back-scattering from the envelope onto the optically thick disk \citep{bast90} or
by multiple scattering in the outer disk regions \citep{whit93}.

The central extinction lane which bisects the nebula is the signature of the disk.
In the case of CB\,26, this was directly demonstrated by spatially resolved mm observations of
the thermal dust emission from the disk which precisely resembles the position, size, shape,
and orientation of the extinction lane (Launhardt \& Sargent 2001; see their Fig. 2).
The disk must be seen almost edge on since none of the NIR lobes seems to be strongly shadowed.
This is consistent with a  preliminary model to the mm dust emission from the disk which suggests an
inclination angle of $\lesssim$\,5\arcdeg.  An independent estimate of the disk inclination can be
derived from the kinematics of the HHO (cf. Section \ref{sec-dis-hh}).
The morphology and colors of the NIR reflection nebula give no solid clue on which side of 
the disk might be tilted toward the observer. 
While the slightly brighter and bluer south-eastern lobe may indicate that this side is 
tilted toward the observer, the presence of the conical extension to the north-west 
(Fig. \ref{fig-j}) and the possible small displacement of the center of illumination with 
respect to the center of the disk (Fig. \ref{fig-k}) point toward the other direction.
However, the displacement is within the range of the astrometric uncertainties and we cannot 
rule out that differences in the appearance of the bipolar lobes are due to varying extinction.
Evidence for non-uniform extinction comes from an additional, more diffuse extinction lane
between the compact north-western lobe and the more diffuse north-west extension of the NIR
nebula which might be due to either an irregular shape of the remnant
envelope or a foreground filament (see Fig. \ref{fig-col}).
The different resolution of the J, H, and Ks images due to the variation of the
seeing precluded the derivation of dust grain properties from the wavelength
dependence of the extinction lane and the observed polarization
\citep{2003ApJ...588..373W}.

We also note that the south-western bright rim of the globule which is close to CB\,26~YSO\,1
coincides with a radio source that is present in 1.6\,GHz and 4.85\,GHz surveys \citep{greg96,cond94}.
The flux densities at these frequencies are 106 and 30\,mJy, respectively, indicating a non-thermal
origin. Thus, the radio source seems to be unrelated to CB\,26.

\subsection{The central illuminating star} \label{sec-dis-star}

Although the central illuminating star is not visible due to the high extinction,
its presence and location could be unambiguously identified by means of the imaging
polarimetry (cf. Sect.\,\ref{sec-res}). Within the errors it is situated at the center
of the mm disk.

The luminosity of CB\,26~YSO\,1 of $\approx 0.5\,L_\sun$, which was derived by integrating the SED over
4\,$\pi$, is only a lower limit to the intrinsic luminosity of the embedded star since the bipolar geometry
causes an anisotropic radiation field \citep{1997A&A...318..879M} where,
due to the edge-on view, only a fraction of the NIR photons are scattered into the
line of sight. However, anisotropic radiation is not expected for the thermal FIR to mm emission, which
contributes 95\%\ to the total luminosity. Thus, even a very high anisotropy correction factor of 10
would increase the bolometric luminosity by only 50\%. 
On the other hand, heating of such an isolated low-mass cloud core by the interstellar radiation field
might not be negligible (e.g., Siebenmorgen, Kruegel, \& Mathis 1992), resulting in an overestimation of 
the intrinsic luminosity of the embedded YSO.
Together with the dynamic mass estimate for the PMS star by \citet{lasa01} of 0.3\,M$_{\sun}$,
a bolometric stellar luminosity of $\le$\,0.5\,L$_\sun$\
yields an upper limit for the age of 10$^6$\,yr based on the solar-metalicity PMS tracks of
\citet{2000A&A...358..593S}.
 From the SED we calculate a ratio $L_{\rm submm}/L_{\rm bol} \approx 2.5\times 10^{-2}$\ for CB26\,YSO1.
According to \cite{andr93} this ratio would be indicative of a Class\,0 protostar
since it is significantly higher than the threshold value $5\times 10^{-3}$.
However, this threshold value was derived from the assumption of spherical symmetry
which does no longer hold for Class\,I YSOs.
According to the morphology and age, CB26~YSO\,1 is clearly a Class\,I object.
The particularly large discrepancy in this case is most likely due to both the 'unusually'
large and massive disk and the extreme edge-on view.
This also demonstrates that a single parameter like the $L_{\rm submm}/L_{\rm bol}$\ ratio
cannot be used to unambigeously classify an individual source, although it may be useful to
analyse (in a statistical sense) a larger sample of objects.

The IRAS non-detection at 12 and 25\,$\mu$m, which is non-typical for a Class\,I source,
is most likely due to the extreme edge-on morphology and the resulting high extinction for the
MIR emission from hot dust in the inner disk.

Since X-ray activity is a sign of low-mass YSOs \citep{feig99}, we searched
the ROSAT archive for a possible X-ray counterpart.
Indeed, there are deep ROSAT images of this area but the superposition with the strong
supernova remnant 1H~0455+518 \citep{pfeff91} makes it difficult to detect
CB\,26~YSO\,1.

\subsection{The Herbig-Haro flow}  \label{sec-dis-hh}

The newly discovered HHO\,494 is well aligned with the symmetry axis of CB\,26~IRS\,1,
suggesting a physical relation between the two. There is no other star-formation
activity in the region which could possibly be responsible for the Herbig-Haro flow.
The neighboring globules CB\,24 and CB\,25 are quiescent and do not harbor any infrared
sources seen in the 2MASS survey or by IRAS. The only other IRAS source in the vicinity
is 04554+5210 to the north-west, at about the same projected distance as 04559+5200.
It is only detected at 100\,\micron\ with a flux of 4.4\,Jy and there is no evidence
for star formation for this source. The absence of other possible driving sources
for the Herbig-Haro flow in the field supports our view that HHO\,494 is due to the
outflow activity from CB\,26~YSO\,1.

The radial velocity of HHO\,494 of $v_{LSR}^{HHO}=-45\pm7$\,km/s is blueshifted
with respect to the systemic radial velocity of the disk $v_{LSR}^{disk}=5.5\,$km/s
\citep{lasa01}. This indicates that the north-western lobe of CB\,26~IRS\,1 is
tilted toward the observer. The assumption of
a maximum disk inclination of 5$\degr$ (cf. Sect.\,\ref{sec-dis-morph}) leads to a velocity estimate of
the Herbig-Haro flow of at least 500\,km/s. Although such a flow velocity seems rather high, it is not
uncommon for Herbig-Haro flows from Class\,I YSOs, e.g., HH\,111 \citep{reip92},
HH\,395 (Stecklum et al., in prep.) and consistent with the absence of a correlation
between source luminosity and flow velocity \citep{2001ARA&A..39..403R}.

However, an independent estimate of the total flow speed can be derived from the detection of HHO\,494
on the POSS-II-F and CCD narrow-band images. The epoch difference between these observations
amounts to 4.4\,yr. The positional coincidence suggests that any proper motion must be smaller
than the pixel size of $\sim$1\arcsec{}. With the assumption of a neglible proper motion of the mm disk,
this yields an upper limit for the tangential velocity component
of 150\,km/s which implies an overall flow velocity of $\lesssim$160\,km/s. Since this estimate of the
tangential velocity is based on direct measurements, we prefer it over the value resulting from the
radiative transfer model. Consequently, the
disk inclination cannot be smaller than 18\degr{} if the jet is perpendicular to the disk, i.e. in the
absence of any precession. When combined with the linear separation of the HHO
from CB\,26~YSO\,1 the upper bound on the tangential velocity yields a lower limit for the dynamic
time-scale of about 1600\,yr.

Ratios of the emission line strengths are a good diagnostic of the excitation conditions,
\citep{raga96}.    
The ratio of the [SII] 6717\AA\ and 6731\AA\ lines is insensitive to the electron temperature,
thus allowing the determination of the electron density \citep{czyz86} which amounts to
$n_e\sim50\,$cm$^{-3}$. An upper limit for $n_e$ of 230\,cm$^{-3}$ is set by the signal-to-noise
ratio of the spectrum. The ratio of the [NII] 6583\AA\
and [OI] 6300\AA\ lines is tracing the ionization fraction and almost independent of pre-shock
density and magnetic field strength \citep{hart94}. The derived value for the ionization fraction
of 0.25$\pm$0.05 implies a
total gas density of about 200\,cm$^{-3}$. This rather low density suggests that the Herbig-Haro
flow is propagating into an environment more similar to the interstellar medium than a molecular
cloud. At such densities the braking of the flow by the ambient medium is presumably not very
efficient which might explain the observed high flow velocity.
From the line ratio of [NII] 6583\AA\ to [OI] 6300\AA\ we also infer the shock velocity of
55$\pm$5\,km/s assuming the low-preshock density case of \citet{hart94}. The comparison of
the shock velocity, i.e.,
the relative velocity of the excited matter in the Herbig-Haro flow, to the flow
speed of $\sim$500\,km/s implies that the HHO\,494 does not represent the terminal
shock.

The presence of the HHO indicates recent outflow activity and ongoing accretion. 
Indeed \citet{lasa01} find indirect signatures of ongoing accretion and a possible
disk wind. However, no large-scale molecular outflow has been detected so far 
which might be due the lack of sufficiently dense molecular gas in and around this
nearly dispersed globule.

No HHO was found for the counter flow within
20\arcmin\ south-east of CB\,26~YSO\,1. This is presumably due to the low density of the ambient
matter which precludes shock formation. The wide-field optical images show this area to be devoid
of any globule remnants, contrary to the region north-west of the source.

\section{Summary and conclusions}  \label{sec-sum}

We report high-resolution NIR observations of a YSO in the
Bok globule CB\,26. Launhardt \& Sargent (2001) obtained subarcsecond-resolution mm
images of this source which they interpreted in terms of a young, large, and massive 
protoplanetary disk of mass $\approx 0.1$\,M$_{\odot}$\ surrounded by a remnant envelope 
of mass $\approx 0.1$\,M$_{\odot}$. From the rotation curve of the disk they deduced a 
mass of the central star of $\approx 0.3$\,M$_{\odot}$. Here we use NIR images and 
polarimetry of the reflection nebula to deduce the morphology of the envelope and the 
location and age of the central illuminating star. We also report the discovery of a 
Herbig-Haro flow from this source and derive further constraints on the morphology and 
evolutionary stage of the YSO-disk-envelope system.
The main results are summarized as follows:

\begin{enumerate}

\item
 The NIR reflection nebula has a bipolar shape and is bisected by an extinction lane 
 due to an edge-on seen circumstellar disk. The two lobes are due to light from the 
 obscured central star scattered at grains in the envelope below and above the disk. 
  The non-uniform extinction by either an irregular shape of the remnant envelope 
 or a foreground filament prevents the derivation of the exact inclination angle from the NIR 
 images.
 
 \item The very small value of the disk inclination suggested by the mm results 
($\lesssim$\,5\degr) is inconsistent with the estimate based on the velocity components of the
Herbig-Haro flow ($\gtrsim$\,18\degr). This discrepancy might be solved by invoking a precession
of the flow, e.g. due to a binary star or other mechanisms \citep{fendt98}. Settling the issue of
the disk inclination requires observations at higher spatial resolution and the application
of 2D/3D radiative transfer models.

\item
 A centro-symmetric polarization pattern and very high polarization degress of 
 $\rm p\sim50\dots90$\%\ indicates the presence of a  single illuminating object.
 From the polarization pattern, its location is derived to be at the center of the 
 extinction lane and disk. The object can be a single star or a close binary.

\item
 From the bolometric luminosity obtained by integrating the SED between $\lambda=$\ 
 0.9\,\micron\ and 3\,mm, the mass of the central star derived from the rotation curve 
 of the disk, and using the PMS evolutionary tracks of \citet{2000A&A...358..593S}, 
 we derive an upper limit for the age of the system of 10$^6$\,yr.

\item
 Evidence for ongoing accretion and outflow activity comes from the discovery of
 HHO\,494 which is well aligned with the symmetry axis of the mm disk. The outflow
 velocity of 160\,km/s implies a dynamical time-scale of 1600\,yr. The gas
 density of 200\,cm$^{-3}$ derived from the electron density and the ionization fraction
 is indicative for a low-density environment of the globule. No counterpart has been
 found for the redshifted flow. 

\item
 CB\,26~YSO\,1 is a classical Class\,I low-mass YSO, i.e., a star-disk system which has
 passed its main accretion phase but is still surrounded by a remnant envelope. 
 Accretion from the envelope onto the disk and from the disk onto the star is still 
 proceeding at low rates and outflow activity has not yet halted. 
 The system is however distinguished by its extreme edge-on view, i.e., the symmetry 
 axis is oriented very close to the plane of the sky. This prevents the detection 
 of NIR and MIR emission from hot dust in the inner part of the disk.

\end{enumerate}

CB\,26\,YSO\,1 resembles very closely both in appearance 
and physical parameters that of the Butterfly star in Taurus 
(IRAS\,04302+2247; Wolf, Padgett, \& Stapelfeldt 2003). 
Both objects are relatively isolated, have fairly large and massive 
circumstellar disks with signatures of grain growth, and are still 
associated with thin remnant envelopes. In contrast to the more evolved, 
smaller, and lower-mass TTS disks, these disks still accrete at low rates 
from the envelope, thus attaining the maximum mass and size during their 
evolution. We suggest that they represent the transition stage from 
embedded protostellar accretion disks to 'naked' protoplanetary disks 
around TTS in an undisturbed environment.

\acknowledgments
We thank R. Lenzen for his support during the MAGIC observations. 
O.F. and B.S. acknowledge travel grants from the Deutsche Forschungsgemeinschaft (
DFG Fi~630/1-1, DFG Ste~605/7-1).
This research has made use of the VizieR data base \citep{ochs00} and the NASA's 
Astrophysics Data System Bibliographic Services.
The Second Palomar Observatory Sky Survey (POSS-II) was made by the California
Institute of Technology with funds from the National Science Foundation, the
National Aeronautics and Space Administration, the National Geographic Society,
the Sloan Foundation, the Samuel Oschin Foundation, and the Eastman Kodak
Corporation.

\clearpage

\begin{deluxetable}{lcccrrl}
\footnotesize
\tablewidth{0pt}
\tablecaption{NIR-Photometry of CB\,26~IRS\,1\label{tab1}}
\tablehead{
\colhead{Filter}&\colhead{Central Wavelength}&\colhead{Aperture}&\colhead{FWHM}&\colhead{F$_\nu$}&\colhead{$\sigma$F$_\nu$}&\colhead{Comments}\cr
&\colhead{[$\mu$m]}&\multicolumn{1}{c}{[\arcsec]} &\colhead{[\arcsec]} &\colhead{[mJy]}&\colhead{[mJy]} &} 
\startdata
 I & 0.90 &24 & 4.0 & 0.062&0.019&\\
 J & 1.25 &12 & 0.9  & 2.2 &0.2& \\
 H & 1.65 &12 & 2.0   & 8.2 &0.8&\\
 Ks & 2.16 &12 & 1.9  & 17.1 &1.7&\\
 L' & 3.77 & 6  & 2.1    & 28.5 &14.0& size  3\farcs5$\times$2\farcs8 at P.A. 130\arcdeg\\
\enddata
\end{deluxetable}

\clearpage

\centerline{\psfig{figure=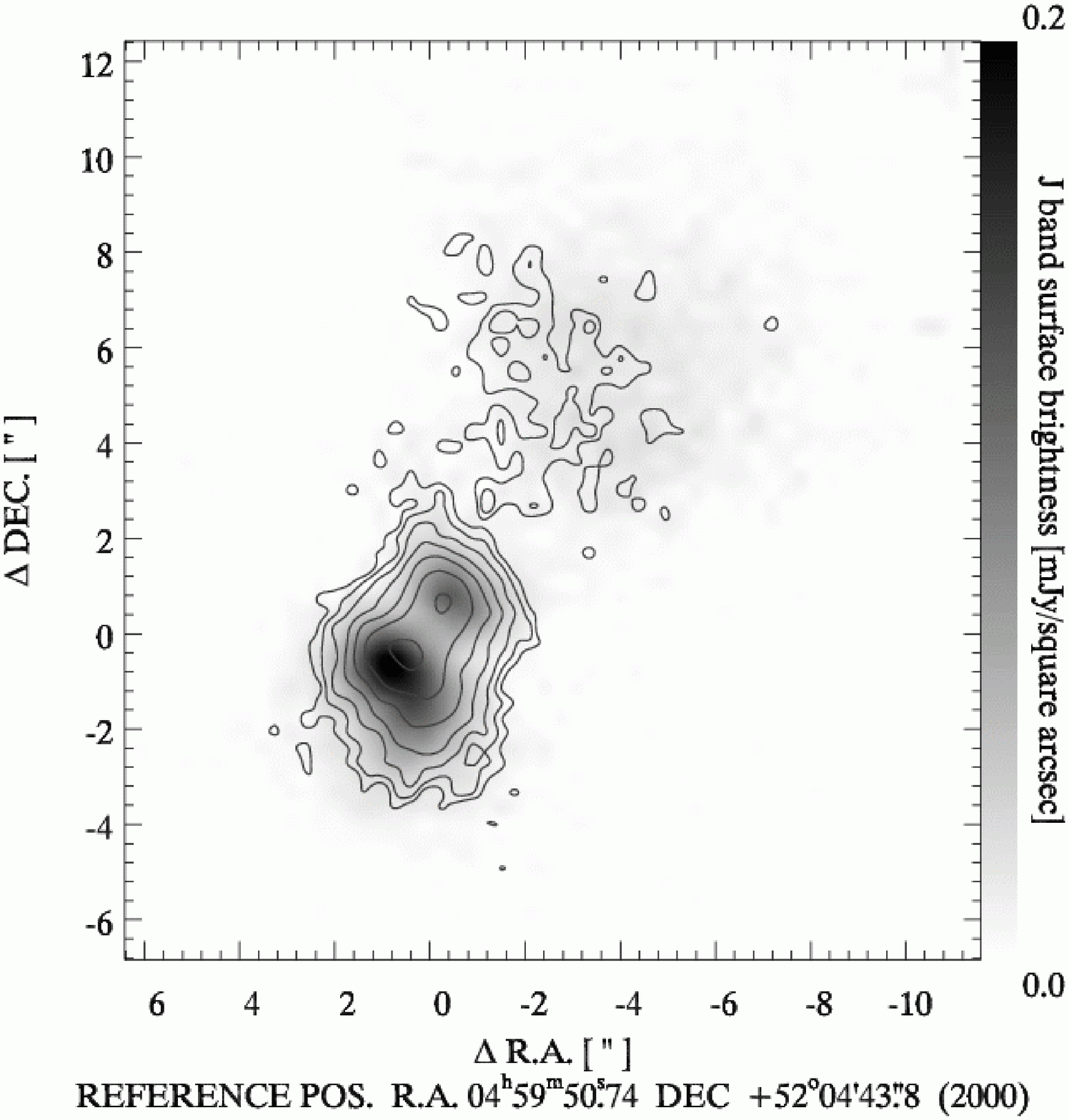,width=3.25in}}
\figcaption{J-band image of CB\,26~IRS\,1 with contours of the K emission (levels are 0.14, 0.22,
0.34, 0.54, 0.86, 1.36, and 2.15 mJy/$\Box$\arcsec.). The shift of the
brightness peaks between J to K is due to the reduced optical depth at longer wavelength.
\label{fig-j}}

\centerline{\psfig{figure=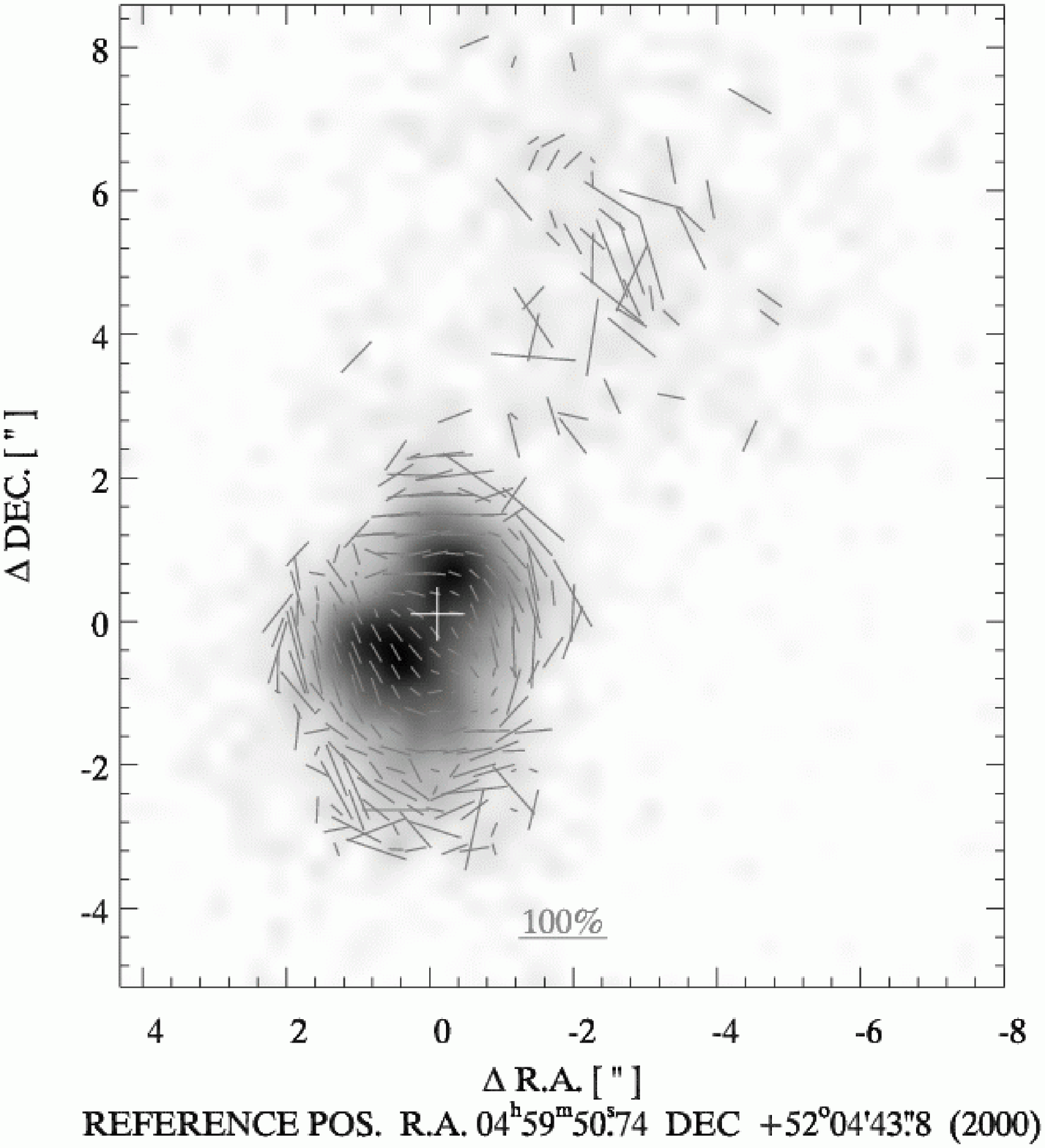,width=3.25in}}
\figcaption{K-band image of CB\,26~IRS\,1 with superimposed vectors of the linear polarization
(the horizontal bar marks $\rm p=100\%$).
The white cross (at the reference position) marks the most likely location of the
illuminator and its size represents the positional uncertainty.
\label{fig-k}}

\centerline{\psfig{figure=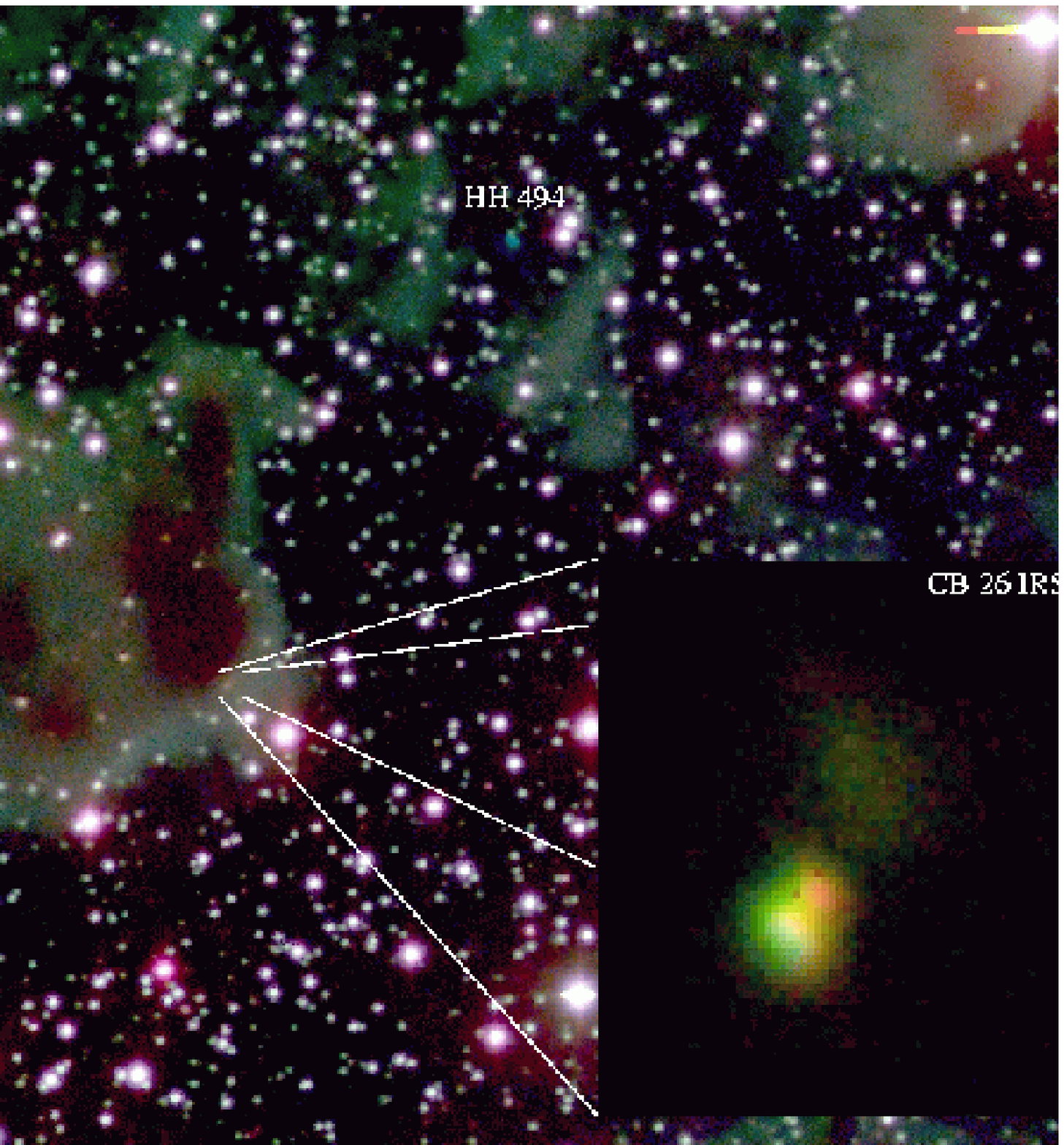,width=3.25in}}
\figcaption{Color representation of the CB\,26 region with IRS\,1 as inset. The optical
image is based on the H$\alpha$ (blue), [SII] (green), and I--band (red) frames while J, H,
and K-band images were used for IRS\,1. The HHO is marked in the optical image. In the
near-infrared color composite, the mm-disk appears as brownish lane at the waist of the
bipolar nebula. 
\label{fig-col}}

\centerline{\psfig{figure=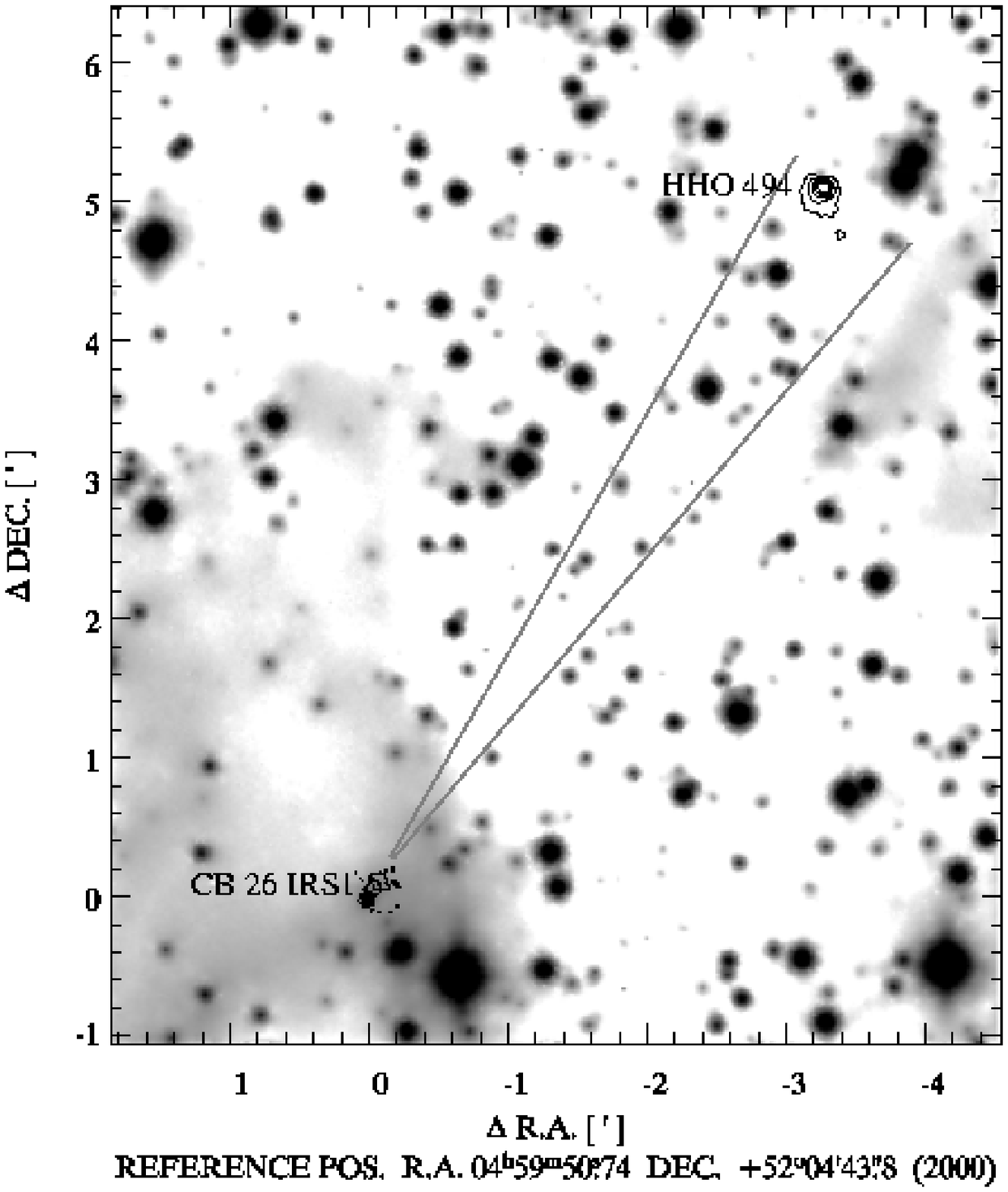,width=3.25in}}
\figcaption{Wide-field I-band image of CB\,26 and its surroundings (gray-scale).
Overlaid are contours of the continuum-subtracted H$\alpha$ image (HHO\,494) 
and of the J-band image of CB\,26~IRS\,1 (located at the reference position).
The gray lines represent the symmetry  axes error cone of CB\,26~IRS\,1 
(see Sect. \ref{sec-res}).
\label{fig-i}}

\centerline{\psfig{figure=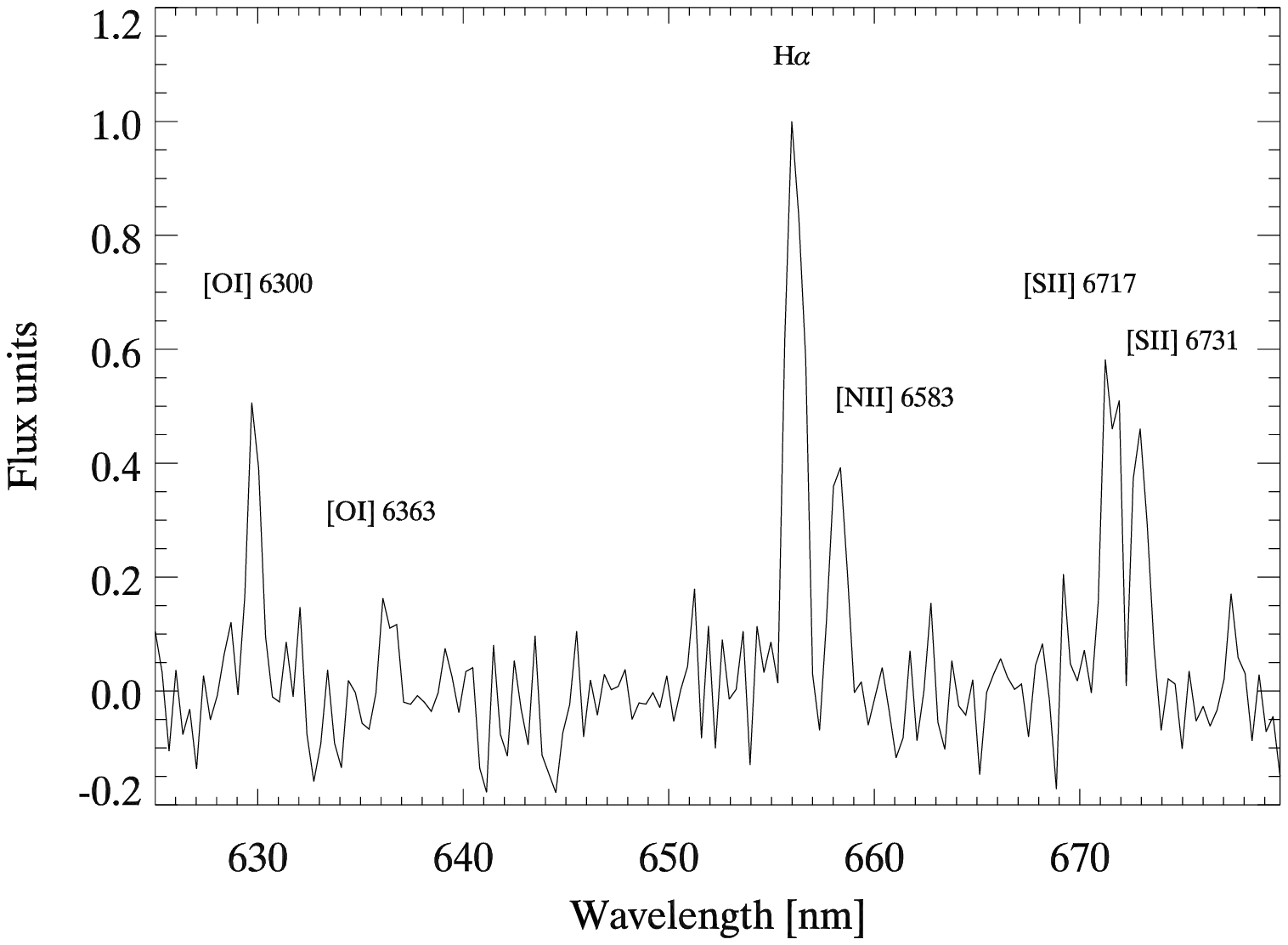,width=3.25in}}
\figcaption{Spectrum of HHO\,494 in the region of the H$\alpha$ line. 
Line fluxes are normalized to that of H$\alpha$. Prominent emission lines are marked.
\label{fig-spec}}

\centerline{\psfig{figure=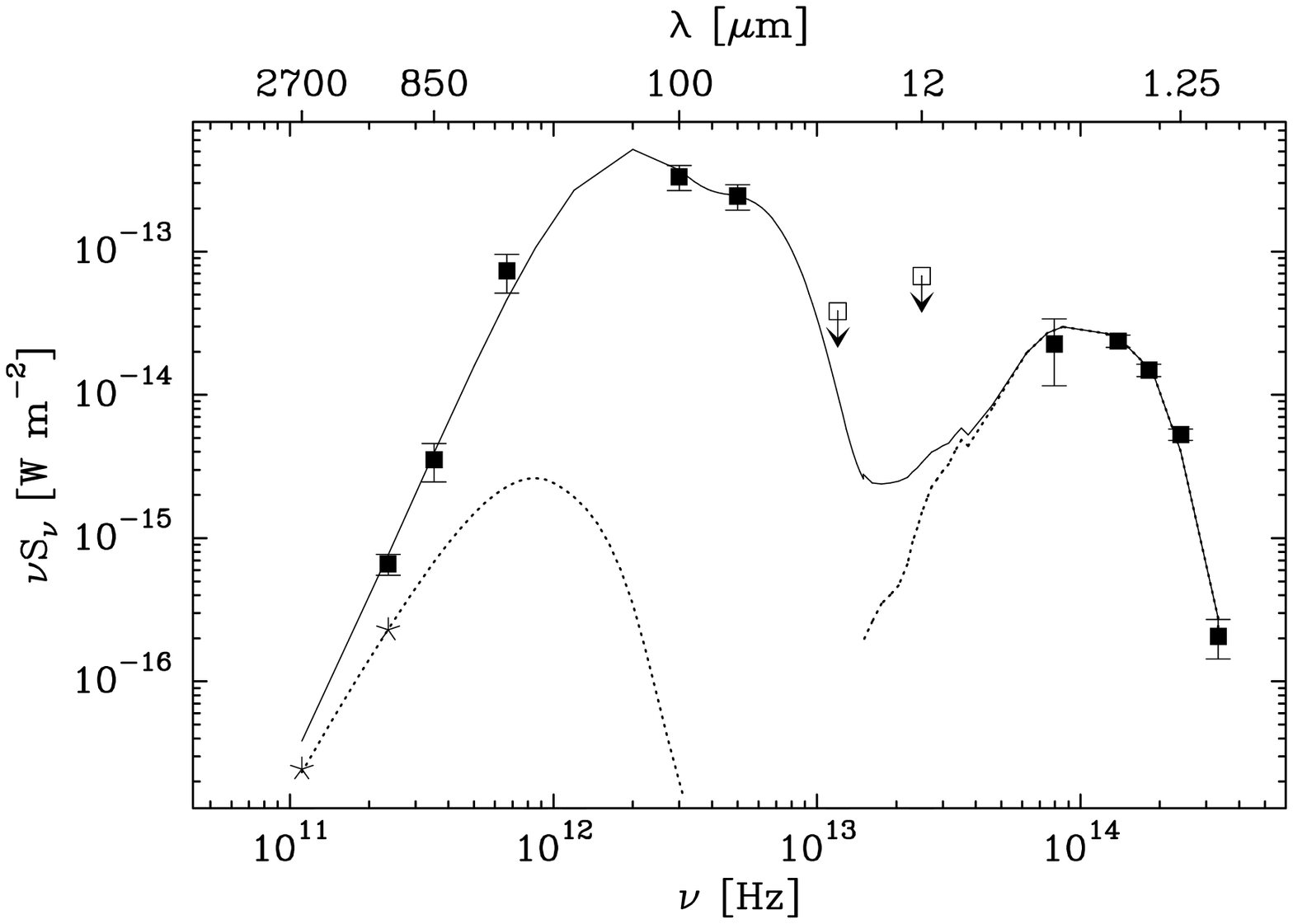,width=3.25in}}
\figcaption{SED of CB\,26~IRS\,1. Full squares represent flux densities from the present paper,
\citet{laun97}, \citet{henn01}, Launhardt et al. (in prep.),
and IRAS PSC. The upper limits of IRAS at 12 and 25\,\micron{}
are marked by open squares. Asterisks represent the submm/mm fluxes for the disk only. The solid
line represents the model,
contributions from the disk and the stellar photosphere are indicated by dotted lines.
\label{fig-sed}}


\begin{thebibliography}{}

\bibitem[Andr\'e et al.(1993)]{andr93}
        Andr\'e, P., Ward-Thompson, D., \& Barsony, M. \ 1993, \apj, 406, 122

\bibitem[Bastien \& M\'enard(1990)]{bast90}
        Bastien, P., \& M\'enard, F. \ 1990, \apj, 364, 232

\bibitem[Bally et al. (1996)]{ball96}
	Bally, J., Devine, D., \& Alten, V. \ 1996, \apj, 473, 921

\bibitem[Bok(1977)]{bok77}
        Bok, B.J. \ 1977, \pasp, 89, 597

\bibitem[Clemens \& Barvainis(1988)]{clba88}
        Clemens, D.P., \& Barvainis, R. \ 1988, \apjs, 68, 257

\bibitem[Clemens et al.(1991)]{clem91}
        Clemens, D.P., Yun, J.L, \& Heyer, M.H. \ 1991, \apjs, 75, 877

\bibitem[Condon et al.(1994)]{cond94} 
        Condon, J.J., Broderick, J.J., Seielstad, G.A., Douglas, K., \& Gregory, P.C.
        \ 1994, \aj, 107, 1829

\bibitem[Czyzak et al.(1998)]{czyz86}
	Czyzak, S.J., Keyes, C.D., \& Aller, L.H. \ 1986, \apjs, 61, 159 

\bibitem[e.g., Els\"asser \& Staude (1978)]{els78}   
	Els\"asser, \& H., Staude, H.~J. 1978, \aap, 70, L3

\bibitem[e.g., Feigelson \& Montmerle(1999)]{feig99}
        Feigelson, E.D., \& Montmerle, Th. \ 1999, \araa, 37, 363

%
%
\bibitem[Fendt \& Zinnecker (1998)]{fendt98}
	Fendt, C., \& Zinnecker, H. \ 1998, \aap, 334, 750

\bibitem[Fischer et al.(1994)]{fisch94}
        Fischer, O., Henning, Th., \& Yorke, H.W. \ 1994, \aap, 284, 187
	
\bibitem[Guibert(1992)]{1992doss.conf..103G} Guibert, J.\ 1992, ASSL
	Vol.~174: Digitised Optical Sky Surveys, 103
	
\bibitem[Hartigan, Morse, \& Raymond(1994)]{hart94}
	Hartigan, P., Morse, J.A., \& Raymond, J. \ 1994, \apj, 436, 125	

\bibitem[Henning et al.(2001)]{henn01}
        Henning, Th., Wolf, S., Launhardt, R., \& Waters, R. \ 2001, \apj, 561, 871 

\bibitem[Herbst et al.(1993)]{herb93}
        Herbst, T.M., Beckwith, S.V.W., Birk, Ch., Hippler, St.
        McCaughrean, M.J., Mannucci, F., \& Wolf, J.
        \ 1993, in: Infrared Detectors \& Instrumentation, SPIE Proc.
        1946, ed. A.M. Fowler, 605

\bibitem[Gregory et al.(1996)]{greg96}
        Gregory P.C., Scott W.K., Douglas K., \& Condon J.J. \ 1996,
        \apjs, 103, 427

\bibitem[Kim et al.(1994)]{kim94}
        Kim, S.-H., Martin, P. G., \& Hendry, P. D. 1994, ApJ, 422, 164

\bibitem[Launhardt(2001)]{lau01}
        Launhardt, R.\ 2001, IAU Symposium, 200, 117

\bibitem[Launhardt \& Henning(1997)]{laun97}
        Launhardt, R., \& Henning, Th. \ 1997, \aap, 326, 329

\bibitem[Launhardt et al.(1998)]{laun98}
        Launhardt, R., Evans, Neal J.II, Wang, Y., Clemens, D.P., Henning, Th.,
        \& Yun, J.L. \ 1998,
        \apjs, 119, 59

\bibitem[Launhardt \& Sargent(2001)]{lasa01}
        Launhardt, R., \& Sargent, A.I. \ 2001, \apj, 562, L173

%
%

\bibitem[Leung(1985)]{leun85}
        Leung, C.M. \ 1985, in: Protostars \& Planets II, eds. D.C. Black \& M.S.
        Matthews, University of Arizona Press, 104

\bibitem[Men'shchikov \& Henning(1997)]{1997A&A...318..879M} Men'shchikov, 
        A.B., \& Henning, Th. \ 1997, \aap, 318, 879 

\bibitem[Monet et al.(1996)]{mone96}
        Monet, D. et al. \ 1996, USNO-SA2.0, (U.S. Naval Observatory, Washington DC)

\bibitem[Ochsenbein et al.(2000)]{ochs00}
        Ochsenbein F., Bauer P., \& Marcout J.\  2000, A\&AS, 143, 221
         
\bibitem[Ossenkopf \& Henning(1994)]{ossk94}
        Ossenkopf, V., \& Henning, Th. \ 1994, A\&A, 291, 943

\bibitem[e.g., Padgett et al.(1999)]{1999AJ....117.1490P} Padgett, D.L., 
        Brandner, W., Stapelfeldt, K.R., Strom, S.E., Terebey, S., \& Koerner, 
        D. \ 1999, \aj, 117, 1490 
        
\bibitem[Pfeffermann et al.(1991)]{pfeff91}
        Pfeffermann, E., Aschenbach, B., \& Predehl, 
        P. \ 1991, \aap, 246, L28

\bibitem[e.g., Raga, B\"ohm, \& Cant\'o(1996)]{raga96}
	Raga, A.C., B\"ohm, K.~H., Cant\'o, J.\ 1996, Rev. Mex. Astron. Astrofis.
	32, 161

\bibitem[Reipurth, Raga, \& Heathcote(1992)]{reip92} Reipurth,
	B., Raga, A.~C., \& Heathcote, S.\ 1992, \apj, 392, 145
	
\bibitem[Reipurth(1999)]{reip99}
        Reipurth, B. \ 1999, A general catalogue of Herbig-Haro objects,
        2. edition, http://casa.colorado.edu/hhcat

\bibitem[Reipurth \& Bally(2001)]{2001ARA&A..39..403R} Reipurth, B.~\&
Bally, J.\ 2001, \araa, 39, 403

\bibitem[Rieke \& Lebofsky(1985)]{riek85}
        Rieke, G. H., Lebofsky, M. J. 1985, ApJ, 330, L33

\bibitem[e.g., Scarrott \& Rolph(1991)]{scar91}
        Scarrott, S.M., \& Rolph, C.D. \ 1991, MNRAS, 249, 131

\bibitem[Siebenmorgen, Kruegel, \& Mathis(1992)]{1992A&A...266..501S} 
	Siebenmorgen, R., Kruegel, E., \& Mathis, J.~S.\ 1992, \aap, 266, 501

\bibitem[Siess et al.(2000)]{2000A&A...358..593S} Siess, L., 
        Dufour, E., \& Forestini, M.\ 2000, \aap, 358, 593 

\bibitem[Whitney \& Hartmann(1993)]{whit93}
        Whitney, B.A., \& Hartmann, L. \ 1993 \apj, 402, 605

\bibitem[cf., Wolf, Padgett, \& Stapelfeldt(2003)]{2003ApJ...588..373W} 
   Wolf, S., Padgett, D.~L., \& Stapelfeldt, K.~R.\ 2003, \apj, 588, 373 

\bibitem[Yun \& Clemens(1994)]{yun94}
        Yun, J.L., \& Clemens, D.P. \ 1994, \apjs, 92, 145

\bibitem[Zinnecker et al.(1999)]{1999A&A...352L..73Z} Zinnecker, H. et al.
        \ 1999, \aap, 352, L73 

\end{thebibliography}
\end{document}